\documentclass[onecolumn,showpacs,preprintnumbers,amsmath,amssymb]{revtex4}
\usepackage{amssymb}
\usepackage{graphicx,color}% Include figure files

\newcommand {\pt} {p_{_{\rm T}}}
\newcommand{\ber}{\begin{eqnarray}}
\newcommand{\eer}{\end{eqnarray}}
\begin{document}

\title{Improved Quark Coalescence for a Multi-Phase Transport Model}
\author{Yuncun He}
\address{Faculty of Physics and Electronic Technology, Hubei University, Wuhan 430062, China}
\address{Department of Physics, East Carolina University, Greenville, NC 27858, USA}
\author{Zi-Wei Lin}
\address{Department of Physics, East Carolina University, Greenville, NC 27858, USA}
\address{Key Laboratory of Quarks and Lepton Physics (MOE) and Institute of Particle Physics,
Central China Normal University, Wuhan 430079, China}
\email{linz@ecu.edu}

\date{\today}

\begin{abstract}
The string melting version of a multi-phase transport model is often applied to high-energy heavy-ion collisions since the dense matter thus formed is expected to be in parton degrees of freedom. In this work we improve its quark coalescence component, which describes the hadronization of the partonic matter to a hadronic matter. 
We removed the previous constraint that forced the 
numbers of mesons, baryons, and antibaryons in an event to be separately conserved
through the quark coalescence process. A quark now could form either a meson or a baryon  depending on the distance to its coalescence partner(s). 
We then compare results from the improved model with the experimental data on 
hadron $dN/dy$, $p_{_{\rm T}}$ spectra, and $v_2$ in heavy-ion collisions from $\sqrt{s_{_{\rm NN}}}=62.4$ GeV to $5.02$ TeV. We show that, besides being able to describe these observables for low-$p_{_{\rm T}}$ pions and kaons, 
the improved model also better describes the low-$p_{_{\rm T}}$ baryon observables in general, especially the baryon $p_{_{\rm T}}$ spectra and antibaryon-to-baryon ratios for multistrange baryons.
\end{abstract}
\pacs{25.75.-q, 25.75.Ld}
\maketitle

\section{Introduction}
A main aim of heavy-ion collisions is to explore the properties of the deconfined Quark-Gluon Plasma (QGP). Large amounts of experimental data such as the hard probes and collective flows at the Relativistic Heavy Ion Collider (RHIC) and the Large Hadron Collider (LHC) have shown that new matter with extremely high temperature and energy density has been created. 
Pb+Pb collisions at $\sqrt{s_{_{\rm NN}}}=5.02$ TeV at the LHC, the highest energy heavy-ion collisions so far, provide information about the evolution and properties of QGP at an even higher energy density. Theoretical simulations of the spacetime evolution of heavy-ion collisions have been performed with hydrodynamic models~\cite{Huovinen:2001cy,Betz:2008ka,Schenke:2010rr,Bozek:2011if}, transport models \cite{Bass:1998ca,Xu:2004mz,Lin:2004en,Cassing:2009vt}, and hybrid models \cite{Petersen:2008dd,Werner:2010aa,Song:2010mg} and then compared against the experimental data on various observables. These comparisons provide valuable information about the properties of the partonic matter and hadronic matter created in these nuclear collisions.

A multi-phase transport (AMPT) model \cite{Lin:2004en} describes heavy-ion collisions by incorporating fluctuating initial conditions, two-body elastic parton scatterings, hadronization, and hadronic interactions. 
The default version of the AMPT model \cite{Zhang:1999bd,Lin:2000cx}, which involves only minijet partons in the parton cascade and uses the Lund string fragmentation for hadronization, 
can well describe the rapidity distributions and transverse momentum spectra of identified particles observed in heavy-ion collisions at SPS and RHIC. However it significantly underestimates the elliptic flow at RHIC. 
Since the matter created in the early stage of high-energy heavy-ion collisions is expected to be in parton degrees of freedom, the string-melting version of the AMPT model \cite{Lin:2001zk}
was constructed, where all the excited strings from a heavy-ion collision are converted into partons and a quark coalescence model is used to describe the hadronization process.
With more partons interacting in the early stage of the QGP evolution, the string-melting AMPT model is able to well describe the anisotropic flows in large \cite{Lin:2001zk,Lin:2004en,Ma:2016fve} as well as small \cite{Bzdak:2014dia} collision systems. However, the string-melting version of the AMPT model could not describe the proton rapidity distributions \cite{Lin:2004en,Zhu:2015voa} 
or the $\pt$ spectra of hadrons including pions \cite{Lin:2004en}.

Recently the string-melting AMPT model, with a new set of key parameters, has been shown \cite{Lin:2014tya} to reasonably reproduce the pion and kaon yields, $\pt$ spectra, and elliptic flows at low $\pt$  in central and semicentral Au+Au collisions at $\sqrt{s_{_{\rm NN}}}=200$ GeV and Pb+Pb collisions at  $\sqrt{s_{_{\rm NN}}}=2.76$ TeV. 
The same version of the model has also been used to predict particle $dN/dy$, $\pt$ spectra, azimuthal anisotropies, and longitudinal correlations in Pb+Pb collisions at  $\sqrt{s_{_{\rm NN}}}=5.02$ TeV \cite{Ma:2016fve}. 
Nevertheless, this string-melting AMPT model still failed to reproduce the $\pt$ spectra and rapidity distributions of protons \cite{Ma:2016fve}. 
For example, it significantly overestimated the proton yield at midrapidity while significantly underestimating the slope of the proton $\pt$ spectra. 
In addition, antibaryon-to-baryon ratios for multistrange baryons such as $\Xi$ and $\Omega$ are well above one for high-energy heavy-ion collisions, contrary to expectations and far from the experimental data. 

In this work, we improve the quark coalescence component in the string-melting version of AMPT and then check the improved AMPT model against the experimental data of heavy-ion collisions at RHIC and LHC energies. 
We describe in Sec.~\ref{coales} the details of the improvement of the quark coalescence process. 
In Sec.~\ref{meson}, we first introduce the parameters used for the improved string-melting AMPT model and then show the model results of pions and kaons in comparison with the experimental data. 
We then show the model results on baryons in Sec.~\ref{baryon} and antiparticle-to-particle ratios in Sec.~\ref{ratio} with comparisons with data. 
After discussions in Sec.~\ref{discuss}, we conclude with Sec.~\ref{conclusions}.

\section{Improvement of quark coalescence model in a multi-phase transport model}
\label{coales}

In the initial condition of the string-melting version of the AMPT model, we fragment the excited strings into hadrons and then decompose each hadron into its constituent quarks before the parton cascade. As a result, almost all energy produced enters the parton cascade for possible scatterings. This is why the string-melting AMPT model has a clear advantage over the default AMPT model in describing the anisotropic flows in ultrarelativistic heavy-ion collisions. After the parton scatterings, a quark coalescence model is used to describe the hadronization process. It combines a quark with a nearby antiquark to form a meson and combines 
three nearby quarks (or antiquarks) into a baryon (or an antibaryon), regardless of the relative momentum among the coalescing partons. 

For the finite number of quarks and antiquarks in an event, the current quark coalescence model in AMPT \cite{Lin:2004en,Lin:2014tya,Ma:2016fve} searches for a meson partner before searching  for baryon or antibaryon partners. 
Specifically, each quark (or antiquark) has its default coalescence partner(s), which is just the one or two constituent parton(s) from the decomposition of the quark's parent hadron. This assignment is possible since currently we only have two-body elastic parton collisions.
Then for any available (i.e., not-yet-coalesced) quark (or antiquark) that originally came from the decomposition of a meson, the quark coalescence model searches all available antiquarks (or quarks) and selects the closest one in distance (in the rest frame of the quark-antiquark system) as the new coalescence partner to form a meson.
After these meson coalescences are all finished, for each remaining quark (or antiquark) 
the model searches all available quarks (or antiquarks) and selects the closest two in distance  as the new coalescence partners to form a baryon (or an antibaryon).
As a result, the total number of baryons in any event after quark coalescence 
is the same as the total number before, so the number of baryons is conserved by quark coalescence. Similarly, the quark coalescence process also conserves the number of antibaryons and the number of mesons in an event.

We now improve the coalescence method by removing the constraint that forced the separate conservation of the numbers of baryons, antibaryons, and mesons through the quark coalescence for each event. Note that the conservation of the number of net-baryons is still automatically satisfied for each event, as well as the conservation of the number of net-strangeness. 
Specifically, for any available quark, the new coalescence model searches all available antiquarks and records the closest one in relative distance (denoted as $d_M$) as the potential coalescence partner to form a meson. The model also searches all available quarks and records the closest one in distance as a potential coalescence partner to form a baryon, and then searches all other available quarks again and records the one that gives the smallest average distance (i.e., the average of the three relative distances among these three quarks in the rest frame of the three-quark system, denoted as $d_B$) as the other potential coalescence partner to form a baryon. 
When both the meson partner and baryon partners are available, 
the quark will form a meson or a baryon according to the following criteria:
\begin{eqnarray}
 d_B < d_M * r_{BM} &:& {\rm form\; a\; baryon;} \nonumber \\ 
{\rm otherwise} &:& {\rm form\; a\; meson,}
\end{eqnarray}
where $r_{BM}$ is the new coalescence parameter, which controls the relative probability of a quark forming a baryon instead of forming a meson. In the limit of $r_{BM} \rightarrow 0$, there would be no antibaryon formation at all while the minimum number of baryons would be formed due to the conservation of the (positive) net-baryon number. On the other hand, in the limit of $r_{BM} \rightarrow \infty$, there would be almost no meson formation (only 0, 1, or 2 mesons would be formed depending on the remainder when dividing the total quark number in the event by three).
The same coalescence procedure is also applied to all antiquarks.
As a result, the new quark coalescence allows a (anti)quark the freedom to form a meson or a (anti)baryon depending on the distance from the coalescence partner(s). This is a more physical picture; for example, if a subvolume of the dense matter is only made of quarks (with a total number being a multiple of three), it would hadronize to only baryons (with no mesons) as we  would expect.

\begin{figure}
\begin{minipage}[t]{0.45\linewidth}
\includegraphics[width=\textwidth]{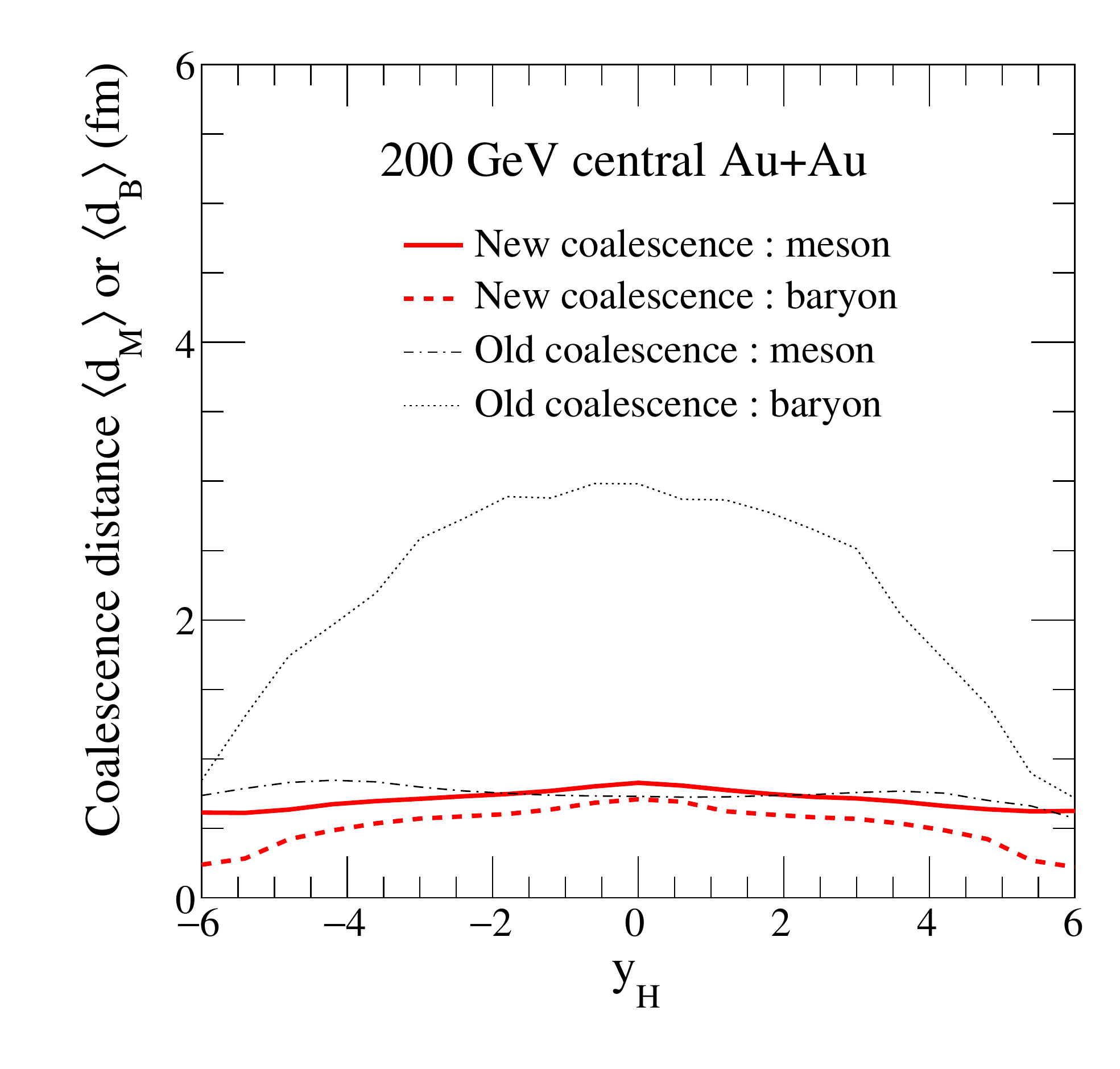}
\end{minipage}
\caption{The average relative distance among coalescing partons in a meson or a (anti)baryon as a function of hadron rapidity from the new (thick curves) and old (thin curves) quark coalescence model for central Au+Au collisions at 200 GeV.
}
\label{fig1-r}
\end{figure}

We take central ($b=0$ fm) Au+Au collisions at $\sqrt{s_{_{\rm NN}}}=200 $ GeV as the example and use the same  parton cross section (1.5 mb) for all the results shown in this section.
Note that the parton phase-space configuration just before quark coalescence is statistically the same for the old and new quark coalescence models (when using the same parton cross section).
Figure~\ref{fig1-r} shows the average relative distance of the quark and antiquark in a primordial meson as well as the average relative distance among the three (anti)quarks in a primordial (anti)baryon as functions of the hadron rapidity ${\rm y_{_H}}$. Here a primordial hadron means a hadron formed directly from quark coalescence (before any hadronic reactions), not a hadron from resonance decays. 
We see that the average relative distance for (anti)baryons from the new quark coalescence (dashed curve) is much lower than that from the old quark coalescence (dotted curve); 
while the average relative distance for mesons is similar to the old model.
This indicates that the new quark coalescence is more efficient, especially for the formation of (anti)baryons, due to the freedom of a parton to form either a meson or a (anti)baryon.

\begin{figure}
\begin{minipage}[t]{0.45\linewidth}
\includegraphics[width=\textwidth]{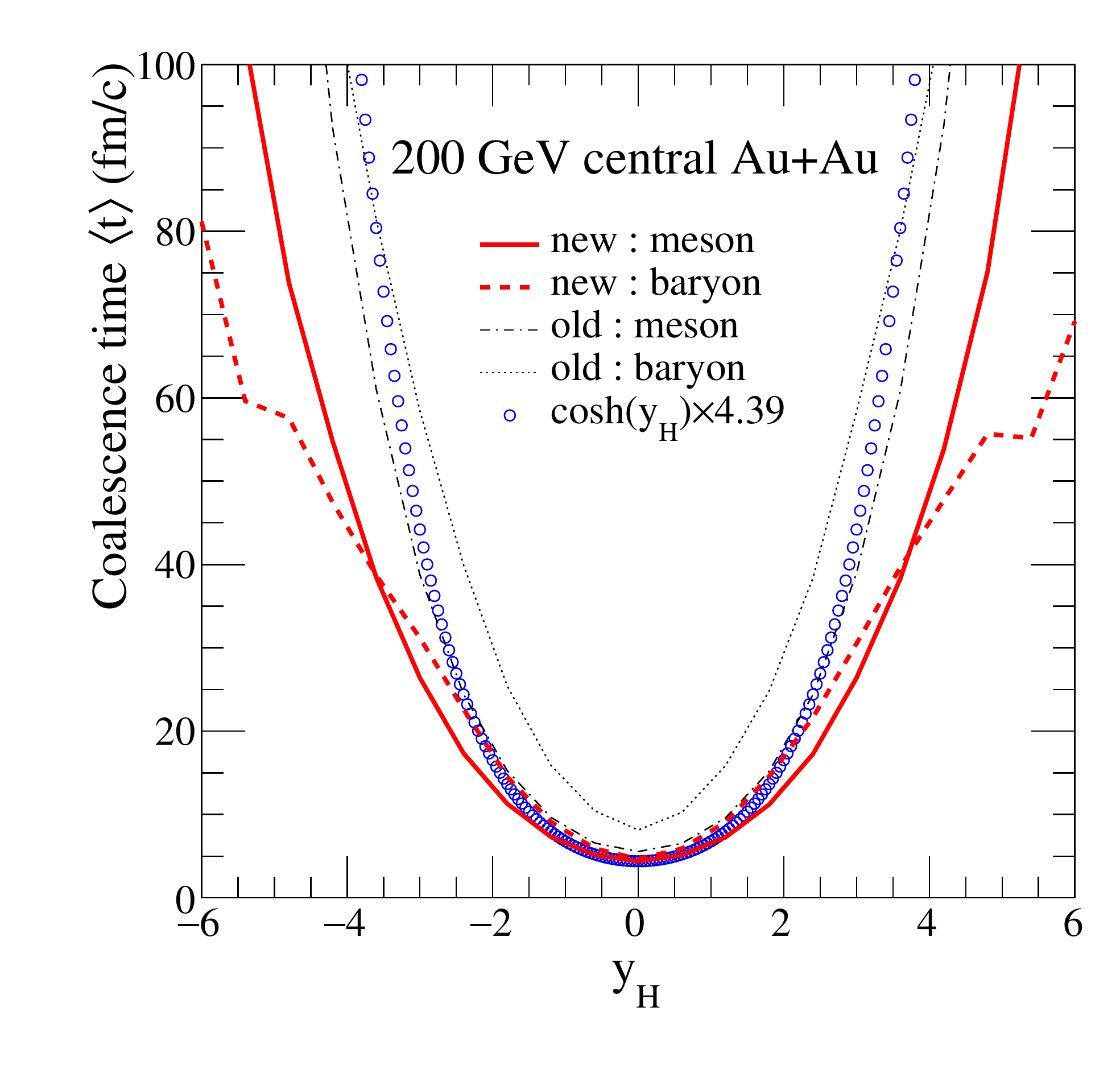}
\end{minipage}
\caption{The average coalescence time of partons in mesons and (anti)baryons as a function of the hadron rapidity from the new (thick curves) and old (thin curves) quark coalescence for central Au+Au collisions at 200 GeV. The circles represent the $\cosh{\rm y_{_H}}$ dependence for comparison.
}
\label{fig2-t}
\end{figure}

Figure~\ref{fig2-t} shows the average coalescence time of partons in mesons and (anti)baryons  as functions of the hadron rapidity ${\rm y_{_H}}$. 
We see that (anti)baryons in the new quark coalescence (dashed curve) are now formed much earlier than before.
The figure also shows that in the old quark coalescence (anti)baryons are formed about 2.6 fm/c later than mesons at midrapidity. 
These are because the old quark coalescence tends to form (anti)baryons late, since it 
searches for meson partners before (anti)baryon partners and a parton will be unavailable for (anti)baryon formation when it is already used for meson formation. 
On the other hand, the new quark coalescence searches for the potential meson partner and (anti)baryon partners concurrently and then determines the hadron type to be formed, making the coalescence process more physical as well as more efficient.
In addition, we see that mesons in the new quark coalescence (solid curve) are also formed earlier than before.
Note that the coalescence time is in the center-of-mass frame, therefore we would expect 
a $\cosh{\rm y_{_H}}$ dependence on rapidity if the dense matter were boost-invariant. 
The circles in Fig.~\ref{fig2-t} represent the curve that is proportional to $\cosh{\rm y_{_H}}$, which qualitatively agrees with our model results.

\begin{figure}
\begin{minipage}[t]{0.45\linewidth}
\includegraphics[width=\textwidth]{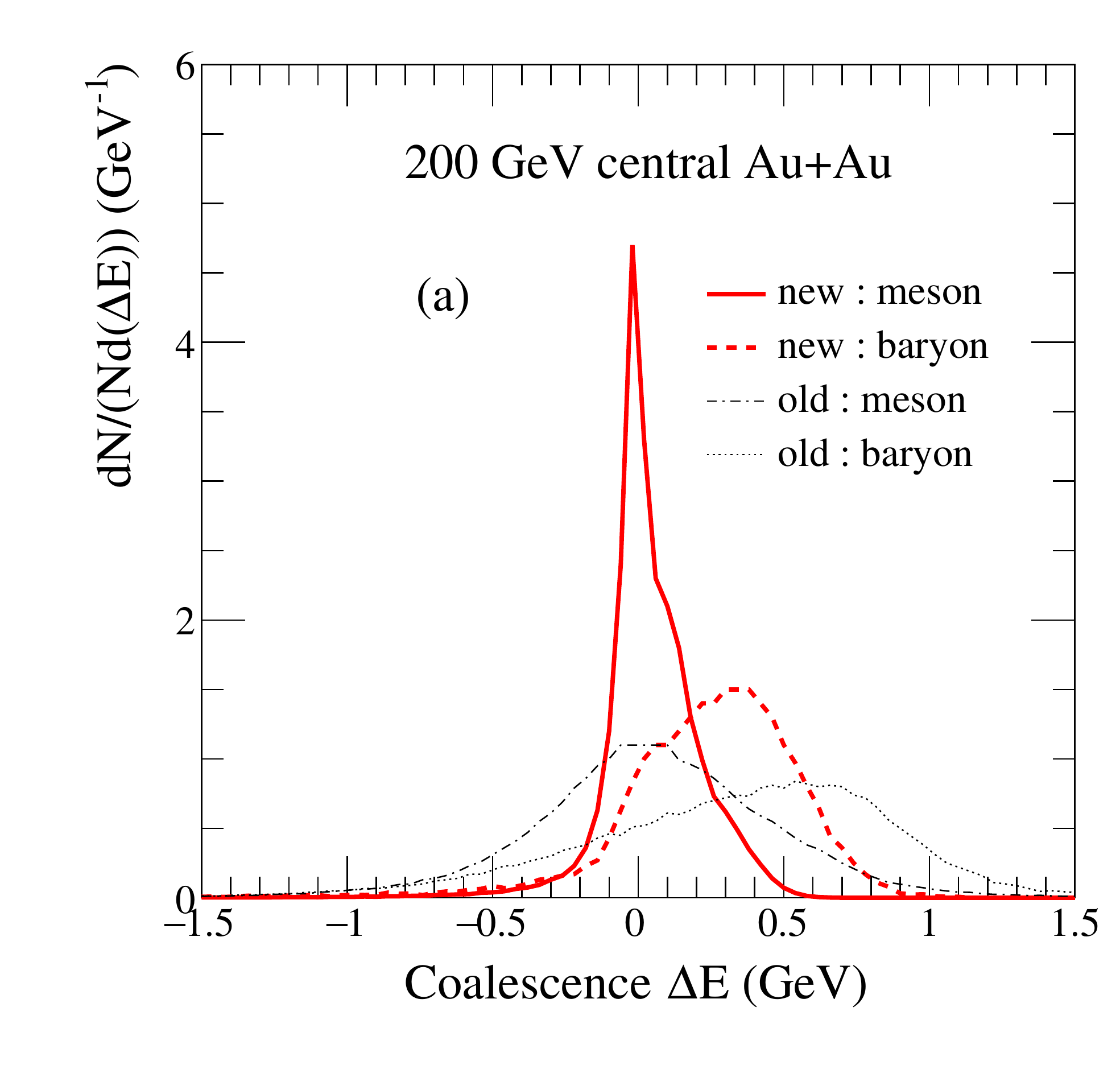}
\end{minipage}
\begin{minipage}[t]{0.45\linewidth}
\includegraphics[width=\textwidth]{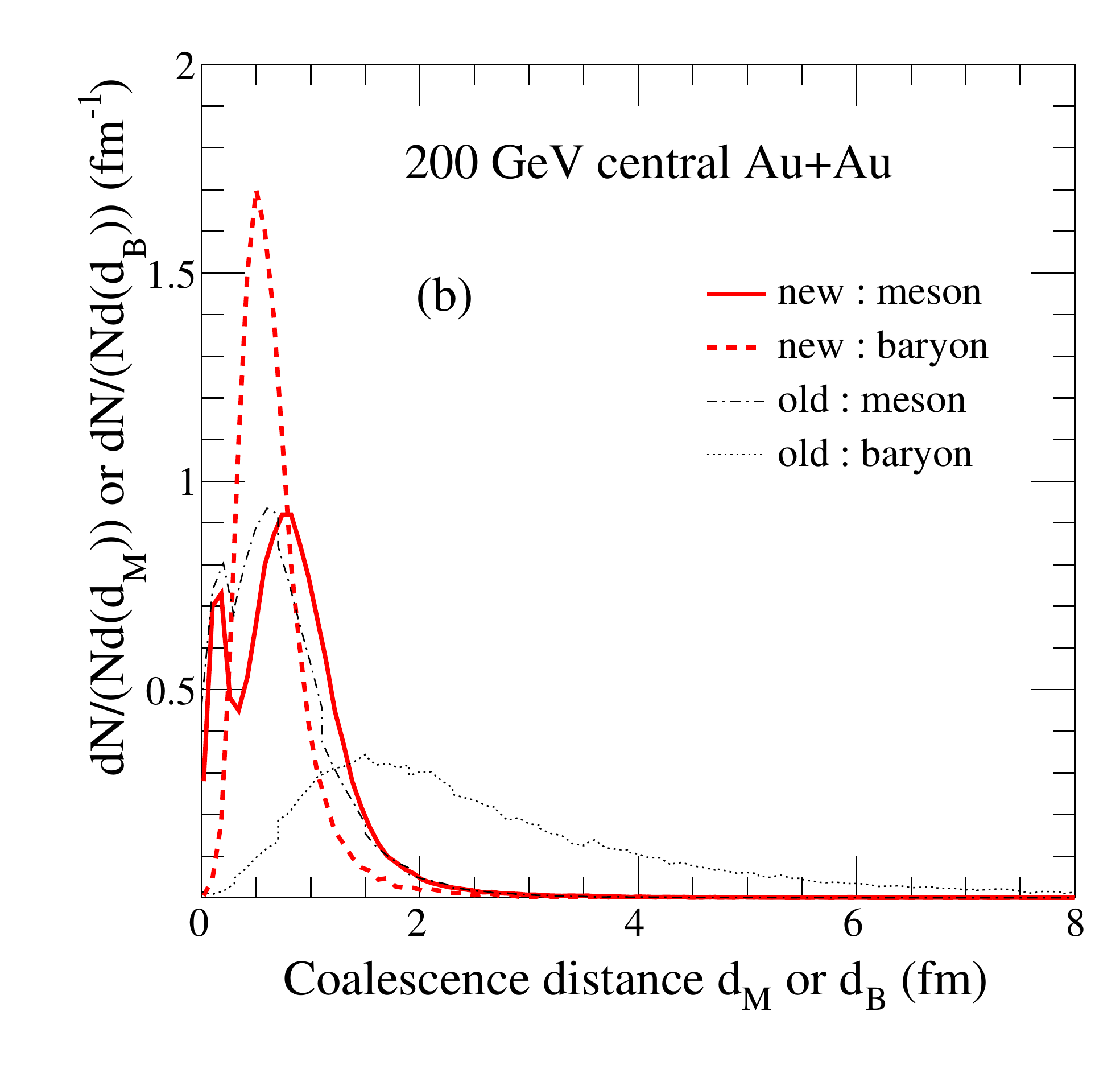}
\end{minipage}
\caption{Normalized distributions of (a) $\Delta E$ (the hadron energy minus the total energy of its coalescing partons) and (b) the relative distance among coalescing partons for the coalescence to mesons and (anti)baryons at midrapidity in central Au+Au collisions at 200 GeV from the new (thick curves) and old (thin curves) quark coalescence.
}
\label{fig3-dE}
\end{figure}

Since the invariant mass of the coalescing quarks for a hadron forms a continuous spectrum rather than a discrete one, the quark coalescence model cannot conserve the four-momentum simultaneously when two or three quarks combine into a hadron. Currently the string-melting AMPT model chooses to conserve the three-momentum during coalescence while violating energy conservation \cite{Lin:2004en}. 
We note that a quark coalescence model that forms resonances with spectral functions \cite{Ravagli:2007xx} or resonance hadrons of finite widths \cite{Bratkovskaya:2011wp} can satisfy the energy-momentum conservation. 
Also note that the AMPT model takes quark masses from the PYTHIA program \cite{Sjostrand:1993yb},  for example, $m_u=5.6$ MeV$/c^2$, $m_d=9.9$ MeV$/c^2$, and $m_s=199$ MeV$/c^2$. 
With $\Delta E$ defined as the difference between the hadron energy and the total energy of its coalescing partons, Fig.~\ref{fig3-dE}(a) shows the $\Delta E$ distribution 
for the coalescence to mesons (solid curve and dot-dashed curve) and (anti)baryons (dashed curve and dotted curve) at midrapidity in central Au+Au collisions at 200 GeV. One would expect a Dirac Delta function $\delta (\Delta E)$ if energy conservation were conserved by quark coalescence,  while the distributions in Fig.~\ref{fig3-dE}(a) have finite widths as a result of the violation of energy conservation. 
However, the widths from the new quark coalescence for both mesons and baryons are narrower than the old results (i.e., the string-melting AMPT results using the old quark coalescence), 
and the peak of the $\Delta E$ distribution for baryons is now closer to the $\Delta  E=0$ position. These indicate that the new quark coalescence performs better in terms of energy conservation. 
Figure~\ref{fig3-dE}(b) shows the distribution of the relative distance $d_M$ (or $d_B$) for the coalescence to mesons (or baryons and antibaryons) at midrapidity in central Au+Au collisions at 200 GeV. We see that the relative distance distribution 
from the new quark coalescence is similar to the old result for mesons but much narrower and closer to zero than the old result for (anti)baryons, consistent with Fig.~\ref{fig1-r}.

\section{Model parameters and the meson spectra}
\label{meson}

For the improved string-melting AMPT model with the new quark coalescence, 
we take the Lund string fragmentation parameters as $a=0.55$ for Au+Au collisions at 200 GeV and $a=0.20$ for Pb+Pb collisions at 2.76 TeV respectively, while $b=0.15$ GeV$^{-2}$. 
These values are the same as those in an early study \cite{Lin:2014tya} except for the Lund $a$ value at 2.76 TeV, which was set to 0.30 before \cite{Lin:2014tya,Ma:2016fve}.
We also use the same strong coupling constant $\alpha_s=0.33$ and keep the upper limit of 0.40 for the relative production of strange to nonstrange quarks from the Lund string fragmentation \cite{Lin:2014tya}. For all the calculations shown in this study, 
the new parameter $r_{BM}$ that controls the relative probability of coalescence of a quark to a baryon  is set to 0.61, the popcorn parameter PARJ(5) that controls the relative percentage of the $B\bar B$ and $B M \bar B$ channels \cite{Lin:2004en} 
is changed to 0.0 (instead of the default value of 1.0), 
and the hadron cascade is terminated at the global time of 200 fm/c. 
The value of the new parameter $r_{BM}$ is chosen in order to reasonably describe the proton $dN/dy$ yields at midrapidity for central Au+Au collisions at 200 GeV and central Pb+Pb collisions at 2.76 TeV (as shown in Fig.~\ref{fig7-pr}).
Note that in this study we set the parton cross section to 1.5 mb, which seems to describe the overall $v_2$ data better than the value of 3 mb used in earlier studies \cite{Lin:2014tya,Ma:2016fve}. Also note that all the ``old'' results (i.e., AMPT results using the old quark coalescence) use the same parameters as those in earlier studies  \cite{Lin:2014tya,Ma:2016fve}, including the 3 mb value for the parton cross section, 
except for the results shown in Sec.~\ref{coales}.

The centrality of AMPT events in this study is determined by the range of impact parameters according to the impact parameter distribution of minimum bias events for a given colliding system at a given energy. For example, for Au+Au collisions at 200 GeV the range of impact parameters 0-3.43 fm represents the 0-5\% centrality while the range 6.76-8.36 fm represents the 20-30\% centrality. For Pb+Pb collisions at 2.76 TeV the range of impact parameters 0-3.74 fm represents the 0-5\% centrality while the range 7.33-8.88 fm represents the 20-30\% centrality.

\begin{figure}
\includegraphics[width=\textwidth]{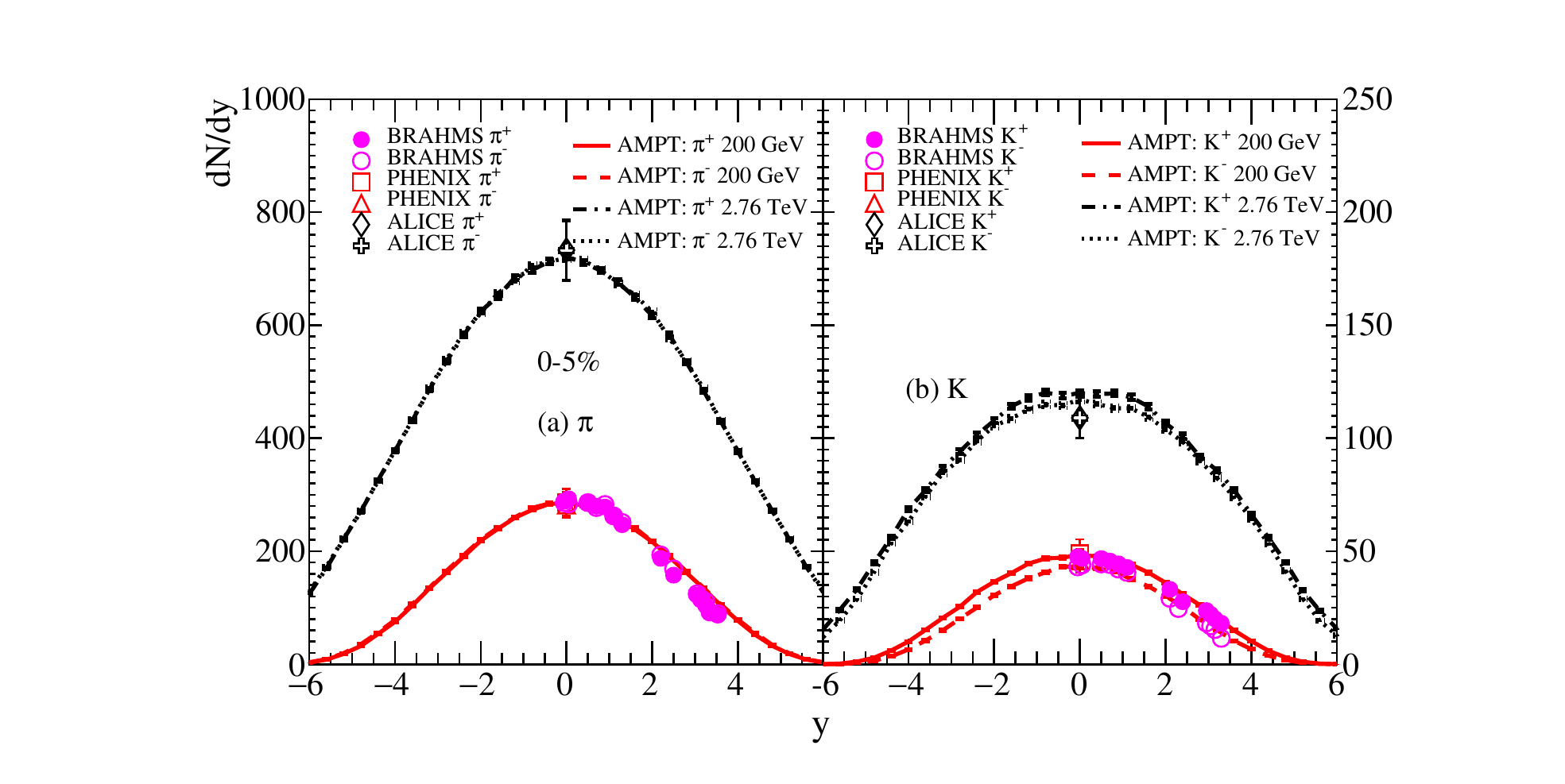}
\caption{$dN/dy$ of (a) charged pions and (b) charged kaons for 0-5\% central Au+Au collisions at 200 GeV and Pb+Pb collisions at 2.76 TeV from the improved string-melting AMPT model (curves) in comparison with the experimental data.}
\label{fig4-dym}
\end{figure}

Figure~\ref{fig4-dym} shows the $dN/dy$ distributions of charged pions and kaons from the improved AMPT model for 0-5\% Au+Au collisions at 200 GeV  in comparison with the experimental data from  PHENIX~\cite{Adler:2003cb} and BRAHMS~\cite{Bearden:2004yx} and for 0-5\% Pb+Pb collisions at 2.76 TeV in comparison with the ALICE data~\cite{Abelev:2013vea}. Note that the BRAHMS pion data have been corrected for $\Lambda$ and $K^0_S$ decays, so they can be directly compared with 
the pion yields from AMPT.
We see that the string-melting AMPT model can reasonably reproduce the $dN/dy$ distributions of charged $\pi$ and $K$. 

\begin{figure}
\includegraphics[width=\textwidth]{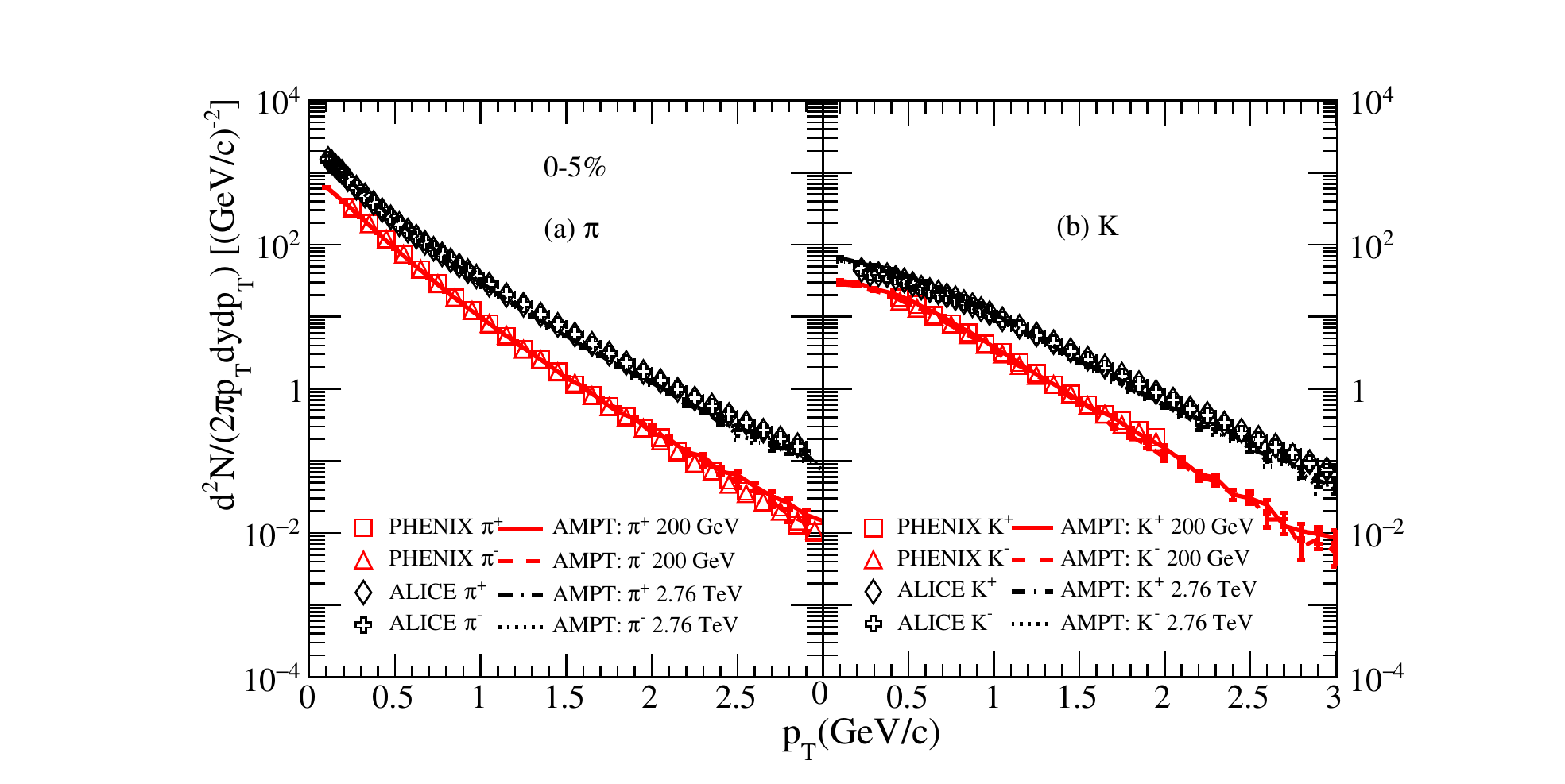}
\caption{The $\pt$ spectra of (a) charged pions and (b) charged kaons at midrapidity for 0-5\% central Au+Au collisions at 200 GeV and Pb+Pb collisions at 2.76 TeV from the improved string-melting AMPT model (curves) in comparison with the experimental data. 
}
\label{fig5-ptm}
\end{figure}

In Fig.~\ref{fig5-ptm} we show the midrapidity $\pt$ spectra of charged pions and kaons from the improved AMPT model for 0-5\% Au+Au collisions at 200 GeV and 0-5\% Pb+Pb collisions at 2.76 TeV in comparison with the experimental data from PHENIX~\cite{Adler:2003cb} and ALICE~\cite{Abelev:2013vea}. 
We see that the AMPT model shows good agreements with the experimental data below $\pt$ of around 2GeV for pions and kaons at both the RHIC and LHC energies. 

\begin{figure}
\includegraphics[width=\textwidth]{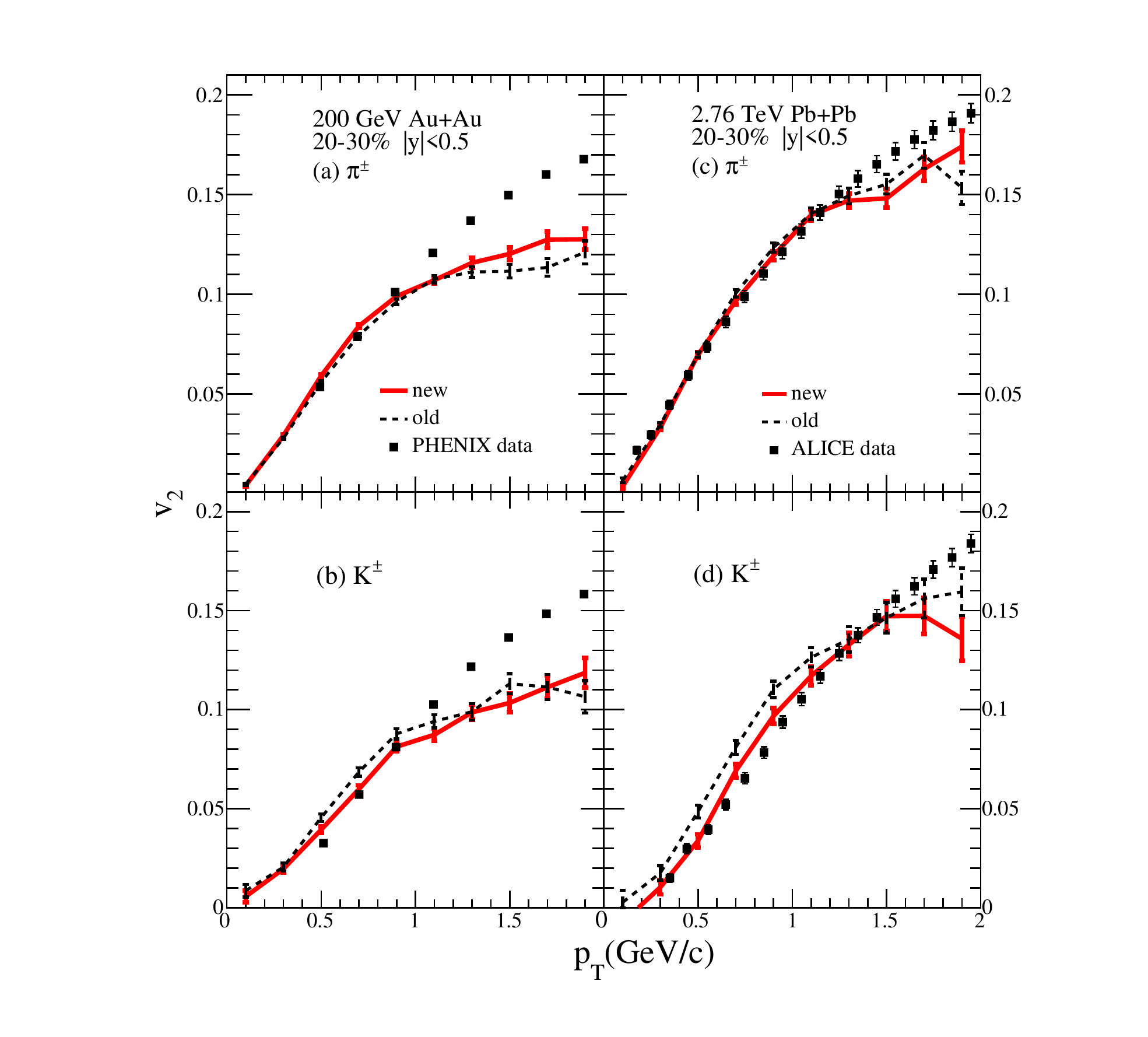}
\caption{Results on $v_2\{\rm EP\}$ of charged pions (upper panels) and charged kaons (lower panels) at midrapidity from the new (solid curves) and old (dashed curves) quark coalescence 
 for 20-30\% central Au+Au collisions at 200 GeV (left panels) and Pb+Pb collisions at 2.76 TeV (right panels) in comparison with the experimental data.
}
\label{fig6-v2m}
\end{figure}

Figure~\ref{fig6-v2m} shows our elliptic flow $v_2\{\rm EP\}$ results of midrapidity charged pions in panels (a) and (c) as well as charged kaons in panels (b) and (d) for 20-30\% central Au+Au collisions at 200 GeV and 20-30\% central Pb+Pb collisions at 2.76 TeV. 
The model results are also compared with the PHENIX \cite{Adare:2014kci} and ALICE  data\cite{Abelev:2014pua}. 
Here the correction factor for the event plane resolution Res$\left\{2\Phi_2\right\}$ is calculated by using charged hadrons in the rapidity ranges of $-2.8<\eta<-1$ and $1<\eta<2.8$ respectively for the two sub-events \cite{ATLAS:2012at}, as done in an early study~\cite{Ma:2016fve}. 
It's seen that the $v_2$ results for charged pions and kaons from the model are generally consistent with the experimental data at low $\pt$.

\section{Baryon spectra}
\label{baryon}

\begin{figure}
\includegraphics[width=\textwidth]{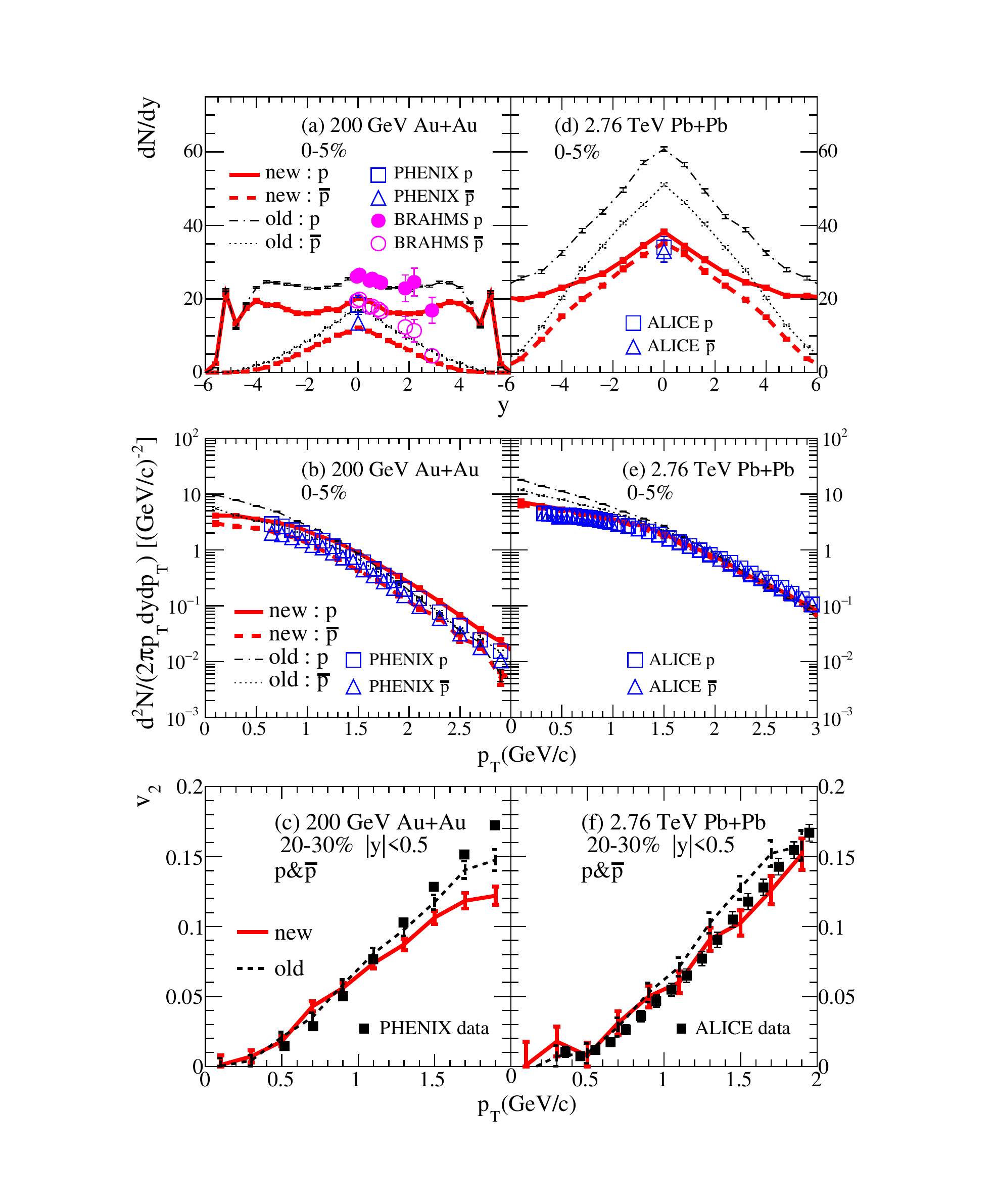}
\caption{$dN/dy$ (upper panels), $\pt$ spectra at midrapidity (middle panels), and $v_2\{\rm EP\}$ at midrapidity (lower panels) of (anti)protons from the new (thick curves) and old (thin curves) quark coalescence for Au+Au collisions at 200 GeV (left panels) and Pb+Pb collisions at 2.76 TeV (right panels) in comparison with the experimental data.  
}
\label{fig7-pr}
\end{figure}

Figure~\ref{fig7-pr} shows the (anti)proton rapidity distributions (upper panels), $\pt$ spectra at midrapidity (middle panels), and $v_2\{\rm EP\}$ at midrapidity (lower panels) for Au+Au collisions at 200 GeV in the left panels and Pb+Pb collisions at 2.76 TeV in the right panels. Both the PHENIX~\cite{Adler:2003cb}  and ALICE~\cite{Abelev:2013vea} data have been corrected for the weak decays of hyperons, thus they can be directly compared with the AMPT results. 
We see that the results from the old quark coalescence (thin curves) significantly overestimate the $dN/dy$ data at midrapidity at both RHIC and LHC energies while also giving too-soft $\pt$ spectra \cite{Zhu:2015voa,Ma:2016fve}.
The results from the new quark coalescence (thick curves) give lower yields and harder $\pt$ spectra and thus better describe the RHIC  and LHC data. 
The proton $v_2$ results are a bit different from the old results but are still mostly consistent with the data at low $\pt$ \cite{Adare:2014kci,Abelev:2014pua}. 

\begin{figure}
\includegraphics[width=\textwidth]{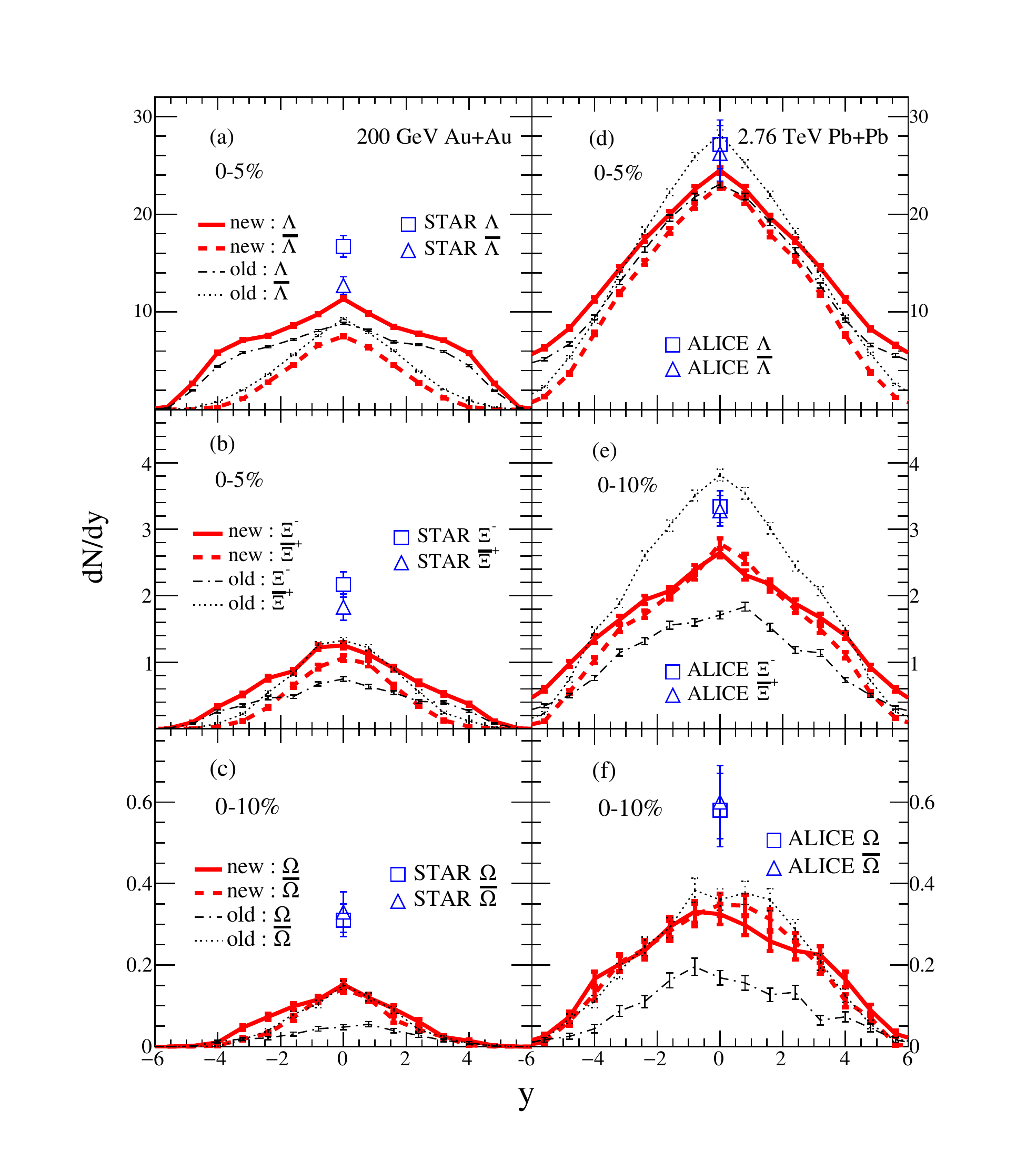}
\caption{$dN/dy$ of $\Lambda$ (upper panels), $\Xi$ (middle panels), and $\Omega$ (lower panels) including the antiparticles from the new (thick curves) and old (thin curves) quark coalescence for central Au+Au collisions at 200 GeV and Pb+Pb collisions at 2.76 TeV 
in comparison with the STAR and ALICE data.
}
\label{fig8-sby}
\end{figure}

Figure~\ref{fig8-sby} shows the $dN/dy$ distributions of strange baryons, including $\Lambda$, $\Xi$, $\Omega$ and their antiparticles, in central Au+Au collisions at 200 GeV (left panels) and  Pb+Pb collisions at 2.76 TeV (right panels) in comparison with the experimental data~\cite{Adams:2006ke,Suire:2002pa,Schuchmann:2015lay,ABELEV:2013zaa}.  
We see that the previous AMPT results (thin curves) for the three types of strange baryons all have more antibaryons than baryons at midrapidity, contrary to the usual expectation for a system with a positive net-baryon number. 
Results from the new quark coalescence in Fig.~\ref{fig8-sby} (thick curves) 
mostly show more baryons than antibaryons as expected. 
We also see that the strange baryon results from both the new and old quark coalescence still underestimate the data, similar to an earlier study based on the default version of AMPT \cite{Pal:2001zw}.
However, the improved AMPT model shows higher yields of strange baryons and is closer to the experimental data than the previous AMPT model.
The underestimation of the strange (anti)baryon yields may be related to the fact that we have not included strangeness production and annihilation processes in the parton cascade of the AMPT model.

\begin{figure}
\includegraphics[width=\textwidth]{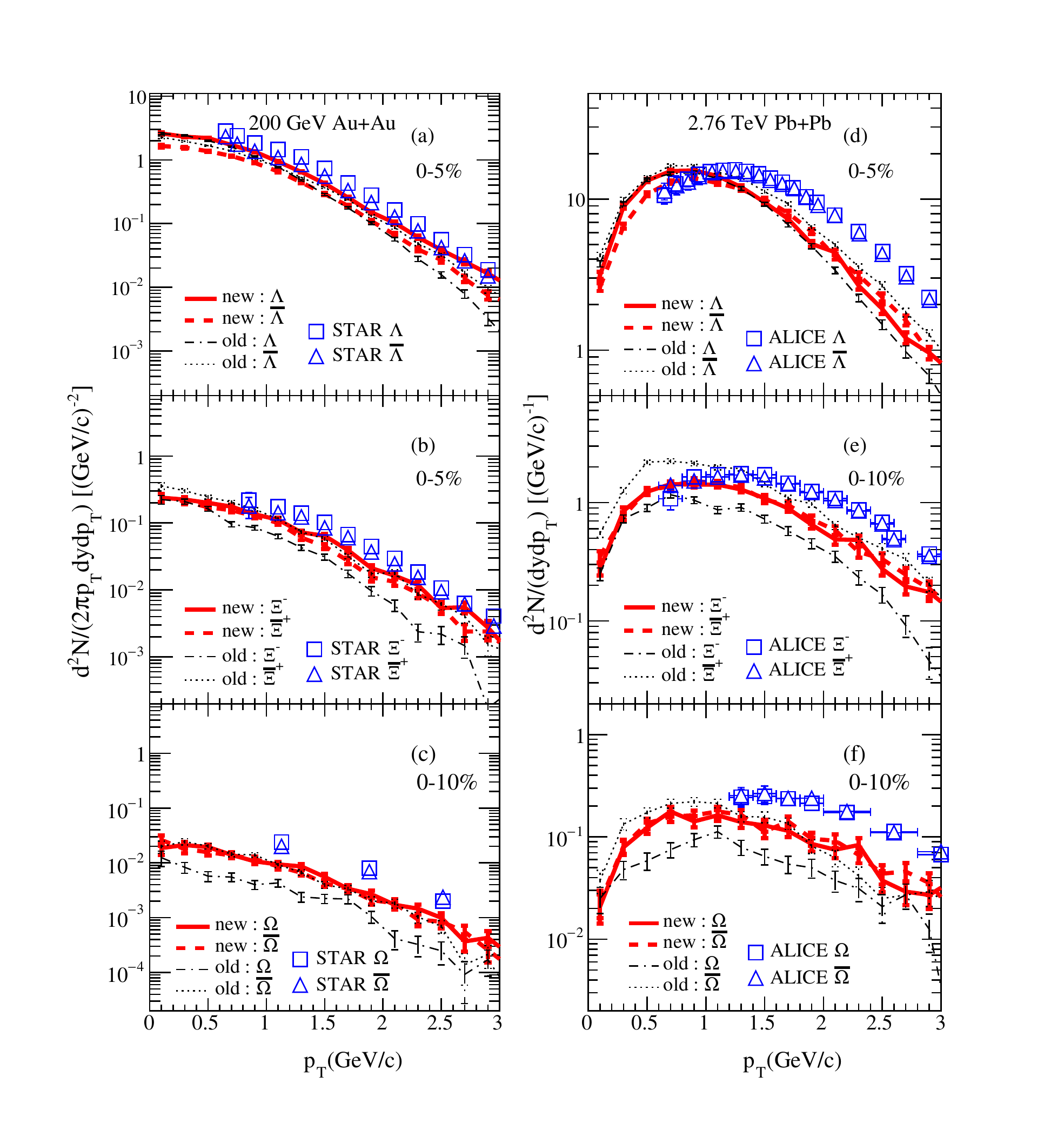}
\caption{The $p_T$ spectra of $\Lambda$ (upper panels), $\Xi$ (middle panels), and $\Omega$ (lower panels) including the antiparticles from the new (thick curves) and old (thin curves) quark coalescence for central Au+Au collisions at 200 GeV and Pb+Pb collisions at 2.76 TeV in comparison with the STAR and ALICE data.   
}
\label{fig9-sbpt}
\end{figure}

Our results on the $\pt$ spectra of strange baryons around midrapidity are shown in  Fig.~\ref{fig9-sbpt}.
The three panels on the left represent respectively the results of $\Lambda$ (0-5\% central), $\Xi$ (0-5\%), and $\Omega$ (0-10\%) in Au+Au collisions at 200 GeV; 
while the three panels on the right represent respectively the results of $\Lambda$ (0-5\% central), $\Xi$ (0-10\%), and $\Omega$ (0-10\%) in Pb+Pb collisions at 2.76 TeV.
We calculated the $\Lambda$ and $\bar \Lambda$ $\pt$ spectra within $|y|<1.0$, the same range as the STAR data. 
To increase the statistics, we calculated the $\pt$ spectra of multistrange baryons (anti)$\Xi$ and (anti)$\Omega$ within the rapidity range of $|y|<1.0$ and $|y|<2.0$, respectively.
Note that, for the AMPT results shown in Figs.\ref{fig10-ratioc}-\ref{fig12-ratios}, 
these rapidity ranges for $\Lambda$, $\Xi$ and $\Omega$ are also used, while 
the AMPT results for protons correspond to midrapidity. 
Also note that the STAR data~\cite{Adams:2006ke,Suire:2002pa} correspond to $|y|<0.75$ for $\Xi$ and $|y|<0.5$ for $\Omega$, while the ALICE  data~\cite{Schuchmann:2015lay,ABELEV:2013zaa} correspond to $|y|<0.5$ for all strange baryons.
We see in Fig.~\ref{fig9-sbpt} that the $\pt$ spectra from the new quark coalescence for strange baryons (thick curves) are harder and compare better with the experimental data, although the $\Lambda$ and $\Xi$ spectra are still softer than the LHC data. 
For strange antibaryons, the $\pt$ spectra from the new quark coalescence have similar slopes as the previous results (thin curves) while the overall magnitudes are mostly lower; however note that the previous yields of strange antibaryons are often much higher than those of strange baryons, contrary to the the data. 

\section{Antiparticle-to-particle ratios}
\label{ratio}

\begin{figure}
\includegraphics[width=\textwidth]{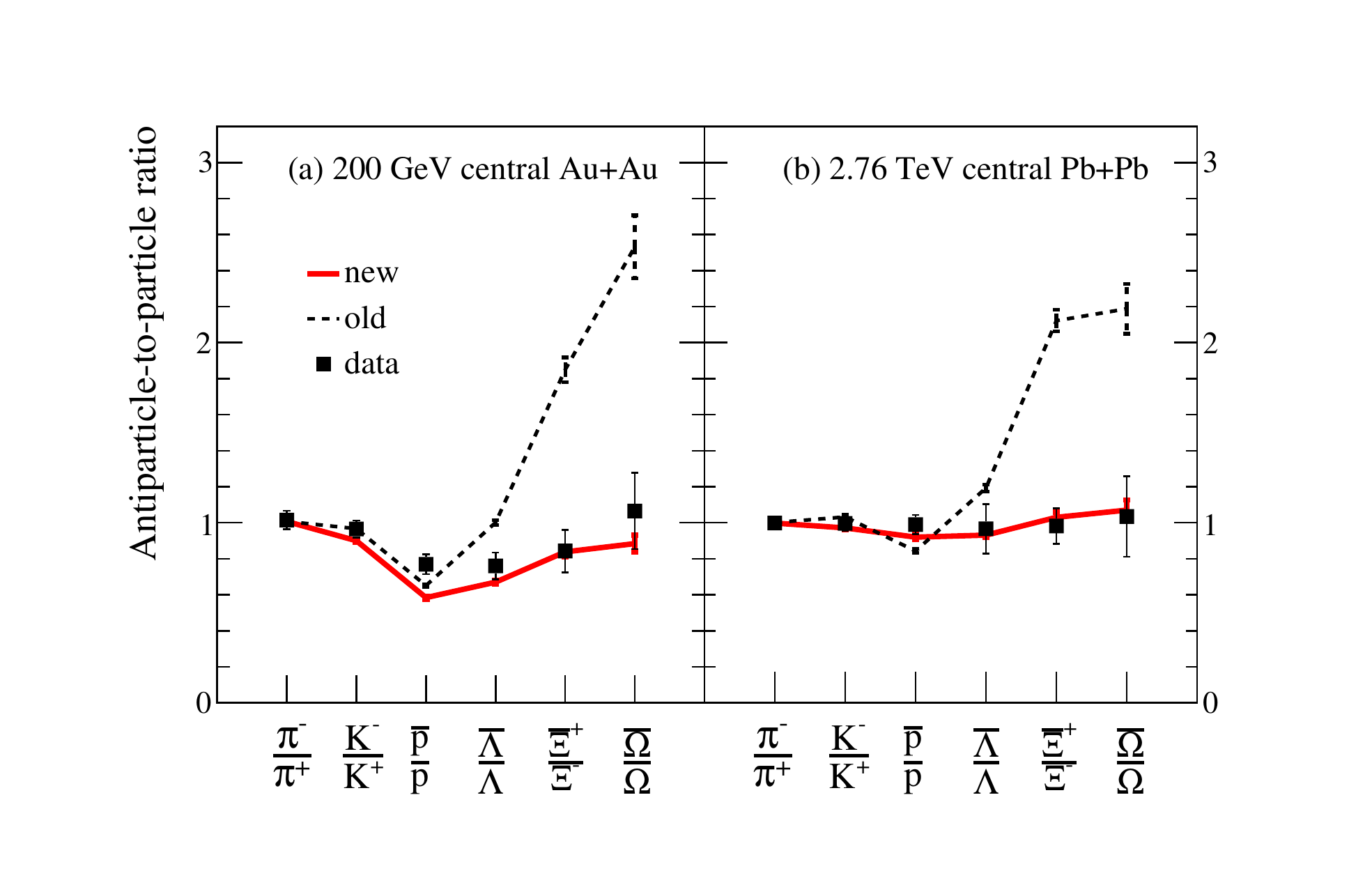}
\caption{Antiparticle-to-particle ratios around midrapidity for central Au+Au collisions at 200 GeV (left panel) and Pb+Pb collisions at 2.76 TeV (right panel) from the new (solid curves) and old (dashed curves) quark coalescence in comparison with the experimental data. 
}
\label{fig10-ratioc}
\end{figure}

Figure~\ref{fig10-ratioc} shows the AMPT results for various antiparticle-to-particle ratios
around midrapidity for central Au+Au collisions at 200 GeV (left panel)~\cite{Suire:2002pa,Adams:2006ke,Abelev:2008ab} and Pb+Pb collisions at 2.76 TeV (right panel)~\cite{Abelev:2013vea,Schuchmann:2015lay,ABELEV:2013zaa} in comparison with the experimental data at midrapidity.
They include the ratios of $\pi^{-}/\pi^{+}$, $K^{-}/K^{+}$, $\bar{p}/p$, $\bar{\Lambda}/\Lambda$,  $\bar{\Xi}^{+}/\Xi^{-}$, and $\bar{\Omega}^{+}/\Omega^{-}$.  
Note that both the data and model results here are for the 0-5\% centrality except that $\Omega$ at 200 GeV and $\Xi$ or $\Omega$ at 2.76 TeV correspond to the 0-10\% centrality.
We see that the results from the new quark coalescence (solid curves) are generally consistent with the experimental data, while results from the old quark coalescence (dashed curves) severely overestimate the ratios for $\Xi$ and $\Omega$. 
We also see that the ratios from the new quark coalescence at 2.76 TeV are closer to the value of one than those at 200 GeV, reflecting the approach to baryon-antibaryon symmetry at midrapidity as the collision energy increases. 
In addition, the antibaryon-to-baryon ratios generally increase with the strangeness content in both the AMPT model and the data. This is consistent with models such as the ALCOR model 
~\cite{Zimanyi:1999py}, which predicts that these ratios are sequentially higher by a multiplicative factor, the $K^{+}/K^{-}$ ratio. 
Since the $K^{+}/K^{-}$ ratio is usually slightly larger than one at high energies, 
we see that our results from the improved quark coalescence agree rather well with this expectation (and with the experimental data).
Note however that the AMPT model underestimates the $\bar{p}/p$ ratio 
in central Au+Au collisions at 200 GeV.

\begin{figure}
\includegraphics[width=\textwidth]{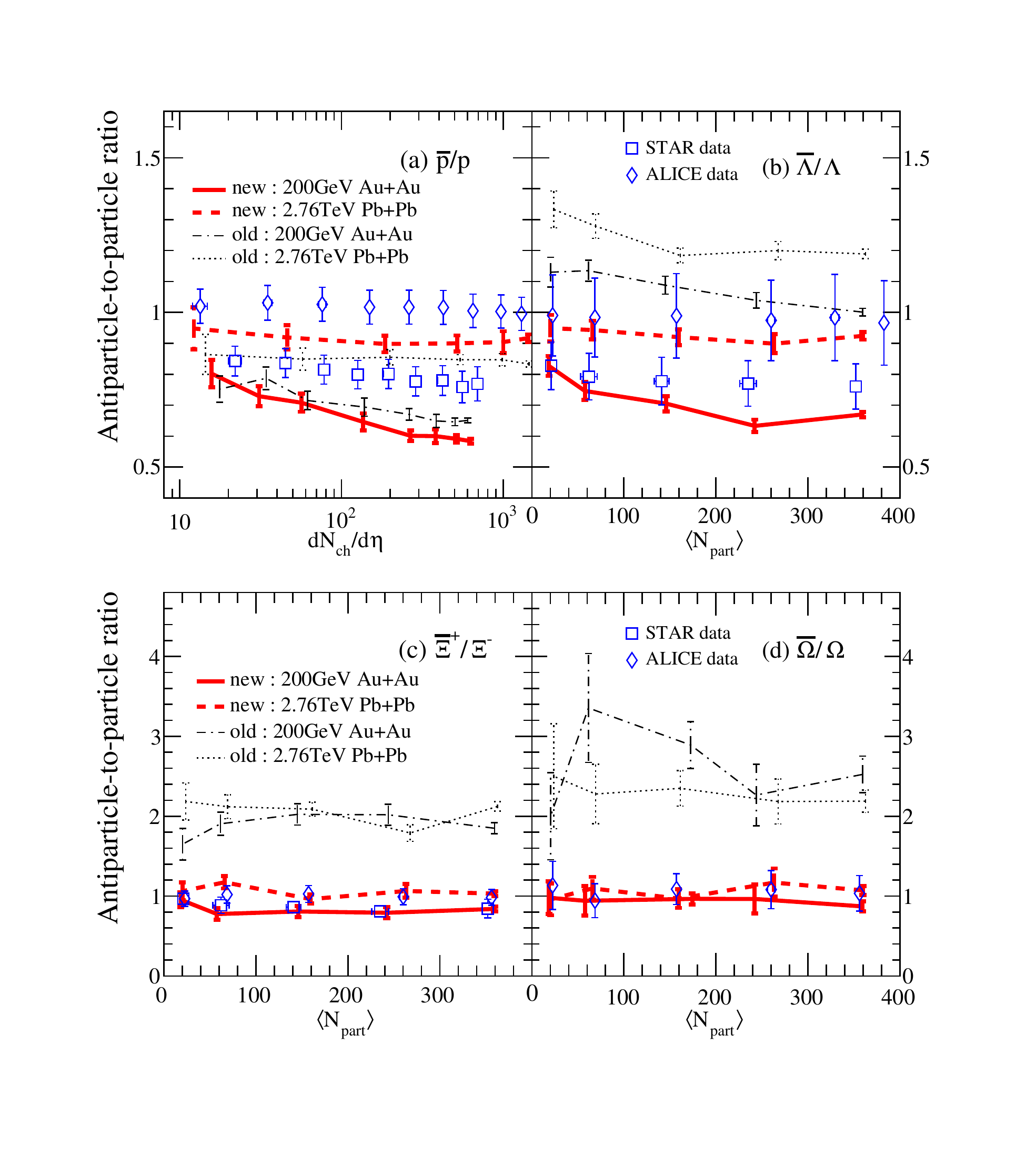}
\caption{
Centrality dependences of antibaryon-to-baryon ratios around midrapidity for (a) protons, 
(b) $\Lambda$, (c) $\Xi$, and (d) $\Omega$ in Au+Au collisions at 200 GeV and Pb+Pb collisions at 2.76 TeV from the new (thick curves) and old (thin curves) quark coalescence in comparison with the experimental data.  
}
\label{fig11-cent}
\end{figure}

Figure~\ref{fig11-cent}  shows the centrality dependence of antibaryon-to-baryon ratios for protons, $\Lambda$, $\Xi$, $\Omega$ around midrapidity in Au+Au collisions at 200 GeV~\cite{Abelev:2008ab,Adams:2006ke}  and  Pb+Pb collisions at 2.76 TeV~\cite{Abelev:2013vea,Schuchmann:2015lay,ABELEV:2013zaa}, where the AMPT results are compared with the STAR and ALICE data for midrapidity ($|y|<0.5$).
We see weak-to-no centrality dependence for these antibaryon-to-baryon ratios, both in the model results and in the experimental data. 
In addition, we see in panels c) and d) that the ratios for $\Xi$ and $\Omega$ 
from  the improved AMPT model (thick curves) 
agree with the experimental data, a significant improvement from the previous AMPT model which gives these ratios well above one. A similar trend is also seen in panel b) 
for $\Lambda$, where the previous AMPT results are much higher than the data. 
However, the improved AMPT model still underestimates the $\bar{p}/p$ ratios at 
both energies and the $\bar{\Lambda}/\Lambda$ ratio at 200 GeV.

\begin{figure}
\includegraphics[width=\textwidth]{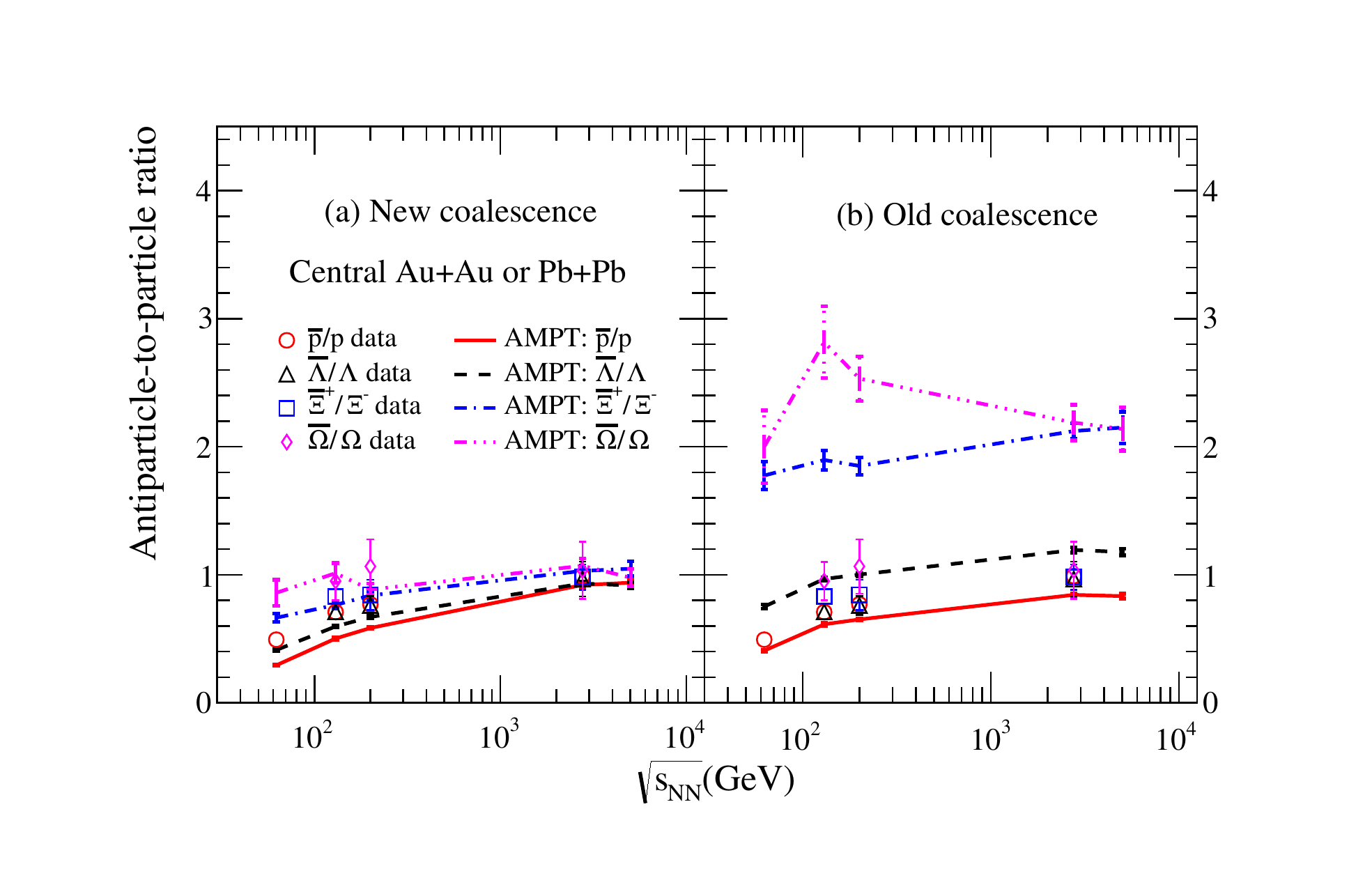}
\caption{
Energy dependences of antibaryon-to-baryon ratios around midrapidity in central Au+Au at RHIC energies and Pb+Pb collisions at LHC energies from the new (left panel) and old (right panel) quark coalescence in comparison with the experimental data. 
}
\label{fig12-ratios}
\end{figure}

Figure~\ref{fig12-ratios} shows the energy dependence of the antibaryon-to-baryon ratios around midrapidity in central Au+Au collisions from 62.4 GeV to 200 GeV and central Pb+Pb collisions at 2.76 TeV and 5.02 TeV.
The Lund string fragmentation parameter $a$ is set to 3.2 at 62.4 GeV and 1.65 at 130 GeV respectively. 
Note that the AMPT results in this figure correspond to the same centrality as the corresponding midrapidity experimental data: 0-10\% central at 62.4 GeV~\cite{Arsene:2009aa}, 0-6\% central for protons and 0-11\% central for strange baryons at 130 GeV~\cite{Abelev:2008ab,Adams:2002pf}, 
0-5\% central for protons, $\Lambda$ or $\Xi$ and 0-10\% central for $\Omega$ at 200 GeV~\cite{Adler:2003cb,Adams:2006ke,Suire:2002pa}, 
0-5\% central for protons or $\Lambda$ and 0-10\% central for $\Xi$ or $\Omega$ at 2.76 TeV~\cite{Abelev:2013vea,Schuchmann:2015lay,ABELEV:2013zaa}. 
Also note that our 5.02 TeV results correspond to 0-5\% central for protons, $\Lambda$ or $\Xi$ and 0-10\% central for $\Omega$.
We see that the results from the new quark coalescence (left panel) are generally more consistent with the experimental data than the old results (right panel), especially for $\Xi$ and $\Omega$ baryons.

\section{Discussions}
\label{discuss}

Our results here have demonstrated that the implementation of quark coalescence in a transport model approach can have significant effects on certain hadronic observables, in this case, the baryon $\pt$ spectra \cite{Zhu:2015voa} and antibaryon-to-baryon ratios \cite{Xu:2016ihu}.
Since the quark coalescence process in our transport model treats the coalescence of individual quarks dynamically, it automatically conserves the total quark number of each flavor in each event, unlike the usual quark coalescence formulation 
\cite{Fries:2003vb,Greco:2003xt,Molnar:2003ff,Lin:2003jy} that is in principle only applicable when the coalescence probability is small (or when the transverse momentum is above a moderate  value).
For the formation of mesons of momentum $p$, the usual quark coalescence \cite{Fries:2003vb,Greco:2003xt} gives the meson momentum distribution as
\ber
E \frac{dN_M(p)}{d^3p} \propto 
\int d^3 q \left|\Psi_p \, (q)\right|^2 f_1(p_1,x) f_2 (p_2,x),
\eer
where $p_1$ and $p_2$ represents the momentum of the quark and antiquark respectively, 
$p =p_1+p_2$, $q = p_1-p_2$, $\Psi_p(q)$ is the meson wave-function, and $f$ is the particle phase-space distribution. This leads to the number-of-constituent-quark scaling relations, e.g., when quarks and antiquarks of different flavors have the same distribution $f_q$, we then have the following scaling relations between the hadron and quark momentum distributions:
\ber
f_M (p_M) \propto f_q^2(p_M/2), f_B (p_B) \propto f_q^3(p_B/3).
\eer
However, it has been shown analytically \cite{Lin:2009tk} that, when the quark number conservation is enforced, the scaling relation is different at low $\pt$. 
For example, in the limit that all partons coalesce and only form mesons, we have a linear scaling relation at low $\pt$:
\ber
f_M (p_M) = f_q (p_M/2)/2^3, 
\label{fb4}
\eer
where the factor $2^3$ is just the normalization factor for the 3-dimensional momentum space. 
Integrating the above meson distribution over the meson three-momentum yields 
the quark number conservation relation $N_M=N_q$, 
where $N_q$ represents the total number of quarks (or antiquarks) just before coalescence.
In general, it is found \cite{Lin:2009tk} that the hadron momentum spectra at low $\pt$  
after quark coalescence depend on the details of the coalescence dynamics such as the time-dependence of the coalescence probabilities of different hadron species. 
Interestingly, the opposite is found \cite{Lin:2009tk} for hadrons above a moderate $\pt$, where the details of the coalescence dynamics do not affect the number-of-constituent-quark scaling relation \cite{Molnar:2003ff,Lin:2003jy}. 
Note that these scaling relations \cite{Fries:2003vb,Greco:2003xt,Molnar:2003ff,Lin:2003jy,Lin:2009tk} are obtained under the assumption that the two or three coalescing partons move along the same direction with no relative momenta. In reality, the coalescing quarks have finite relative momenta and thus  finite opening angles among them \cite{Li:2016ubw}.

Based on the above discussion, it is not surprising that the order of implementing the quark coalescence in the AMPT model affects the hadron spectra at low $\pt$. 
Earlier we have found that simply reversing the coalescence order (i.e., searching for baryon or antibaryon partners before searching for meson partners) in the AMPT model 
significantly decreased the antibaryon-to-baryon ratios for strange baryons; 
this reversed ordering was used to study light (anti-)nuclei productions \cite{Zhu:2015voa} although it still failed to describe the proton and antiproton $\pt$ spectra in Pb+Pb collisions
at 2.76 TeV. 
In this study, the new quark coalescence has no preference (or order) for meson or baryon formations; instead a quark is free to form either a meson or a baryon depending on the distance to the potential coalescence partner(s). 
We have seen that this has led to significant improvements on the proton and antiproton $\pt$ spectra as well as the antibaryon-to-baryon ratios for strange baryons.
Furthermore, since the improved model no longer forces the conservation of the baryon number (or the antibaryon number) of each event through the quark coalescence process, 
it allows the fluctuation of the baryon number (and correspondingly the fluctuation of the antibaryon number) of each event through quark coalescence; therefore it is also suitable for future studies of the effects of phase transition on baryon number fluctuations.

With an improved quark coalescence, 
the string-melting AMPT model gives similar results for the $dN/dy$ yields and $\pt$ spectra of charged pions and kaons as well as $v_2$ of pions, kaons, and (anti)protons 
as the previous string-melting AMPT model \cite{Lin:2014tya,Ma:2016fve}. 
The model results for these observables are mostly consistent with the experimental data. 
In addition, the improved model gives lower proton and antiproton yields with harder $\pt$ spectra, which now reasonably reproduce the data. 
The improved model also leads to higher yields and harder $\pt$ spectra
of strange baryons, with better agreement with the data. Although the improved model gives lower strange antibaryon yields than the previous model, the strange antibaryon-to-baryon  ratios are now below or close to one and are mostly consistent with the experimental data, while these ratios from the previous model are often well above one and incompatible with the data. Therefore, we conclude that the string-melting AMPT model with the new quark coalescence provides a better overall description of the bulk matter in high-energy heavy-ion collisions.

Despite the more physical nature of the new quark coalescence, there are still some problems; for example, the antiparticle-to-particle ratios for protons and $\Lambda$ at RHIC energies underestimate the experimental data. This indicates that future work is necessary to further improve the description of the bulk matter. For example, the incorporation of up-to-date parton structure functions of nuclei and further developments of the dynamical parton recombination process \cite{Lin:2011zzg,Lin:2014uwa} should make a multi-phase transport model more reliable.

\section{Conclusions}
\label{conclusions}

In the string-melting version of a multi-phase transport model, 
the initial matter formed right after the high-energy heavy-ion collision 
is considered to be in parton degrees of freedom, 
and the subsequent hadronization of the partonic matter to a hadronic matter is modeled by quark coalescence. In this work we improve the quark coalescence component of the AMPT model. 
In particular, we removed the previous constraint that forced the quark coalescence process to conserve the numbers of mesons, baryons, and antibaryons in an event separately, where only the conservation of the net-baryon number needs to be required.
The new quark coalescence now allows a quark to form either a meson or a baryon, depending on the distance to its coalescence partner(s). 
We then compare results from the improved AMPT model with the experimental data in heavy-ion collisions at RHIC and LHC energies from $\sqrt{s_{_{\rm NN}}}=62.4$ GeV to $5.02$ TeV, including the centrality and energy dependences of antibaryon-to-baryon ratios. 
We show that, besides being able to describe the $dN/dy$ yields, $\pt$ spectra, and
elliptic flows of pions and kaons at low $\pt$, the improved model also 
better describes the baryon observables in general, especially the $\pt$ spectra of baryons and antibaryon-to-baryon ratios for $\Xi$ and $\Omega$. 
The string-melting AMPT model with the new quark coalescence thus provides a better overall description of the bulk matter in high-energy heavy-ion collisions.

\section*{Acknowledgments}
We thank Jin-Hui Chen and Huan-Zhong Huang for discussions. 
This work was supported in part by the NSFC of China under grants No. 11628508 (ZWL), No. 11547016 and No. 11447190 (YH). YH also acknowledges the financial support from the China Scholarship Council.

\end{document}